\documentclass[10pt,aps,longbibliography,onecolumn,superscriptaddress]{revtex4-2}

\usepackage[T1]{fontenc}
\usepackage[utf8]{inputenc}
\usepackage[english]{babel}

\usepackage{amsfonts,amsbsy,amssymb,amsmath}
\usepackage{graphicx,float}
\usepackage{multirow}

\usepackage[linktoc=all,colorlinks=true,citecolor=blue,linkcolor=blue]{hyperref}

\graphicspath{{figures/}}

\begin{document}

\title{Reactive Islands for Three Degrees-of-Freedom Hamiltonian Systems}

\author{Vladim{\'\i}r Kraj{\v{n}}{\'a}k}
\email{v.krajnak@bristol.ac.uk, corresponding author}
\affiliation{School of Mathematics, University of Bristol, \\ Fry Building, Woodland Road, Bristol, BS8 1UG, United Kingdom.}

\author{V\'ictor J. Garc\'ia-Garrido}
\email{vjose.garcia@uah.es}
\affiliation{Departamento de F\'isica y Matem\'aticas, Universidad de Alcal\'a, \\ Madrid, 28871, Spain.}

\author{Stephen Wiggins}
\email{s.wiggins@bristol.ac.uk}
\affiliation{School of Mathematics, University of Bristol, \\ Fry Building, Woodland Road, Bristol, BS8 1UG, United Kingdom.}

\begin{abstract}

We develop the geometrical, analytical, and computational framework for reactive island theory for three degrees-of-freedom time-independent Hamiltonian systems. In this setting, the dynamics occurs in a 5-dimensional energy surface in phase space and is governed by four-dimensional stable and unstable manifolds of a three-dimensional normally hyperbolic invariant sphere. The stable and unstable manifolds have the geometrical structure of spherinders and we provide the means to investigate the ways in which these spherinders and their intersections determine the dynamical evolution of trajectories. This geometrical picture is realized through the computational technique of Lagrangian descriptors. In a set of trajectories, Lagrangian descriptors allow us to identify the ones closest to a stable or unstable manifold. Using an approximation of the manifold on a surface of section we are able to calculate the flux between two regions of the energy surface.

\end{abstract}

\maketitle



\section{Introduction}
\label{sec:intro}

The deterministic evolution of a system between two stable states through an intermediate unstable state is a fundamental setting for an important form of dynamical evolution that informs our way of thinking of ``transition phenomena’’. This dynamical situation is common to many diverse fields in science and engineering, such as celestial mechanics\cite{koon2000heteroclinic, gomez2001invariant}, structural mechanics\cite{zhong2018tube, ross2018experimental, zhong2020geometry}, and chemical reaction dynamics. This setting is natural for studies in the latter area and will be the focus of the applications that we consider. 

The mathematical setting that we will consider is that of two and three degrees-of-freedom (DoF) autonomous Hamiltonian systems (hence, four and six dimensional phase spaces, respectively). The Hamiltonians have the form of the sum of potential  energy and kinetic energy, where the potential energy is a function of the position variables (``configuration space’’) and the kinetic energy is a function of the momentum variables (variables conjugate to the position variables). The topography of the potential energy function, or ``potential energy surface’’ (PES) plays an important role in the reaction problem that we study. Potential wells (i.e. minima of the PES) are identified with stable states, or ``reactants’’ and ``products’’, and they are  separated by saddle points on the PES. There are different types of saddle points characterized by their index \cite{Agaoglou2019}. In this paper we will be concerned with index-one saddle points as this is the most common saddle point considered in this type of transition phenomena. Trajectories passing between potential wells are dynamical objects that are most naturally studied in phase space, i.e., the space of configuration and momentum variables. The situation is very different for two and three DoF systems, and we will describe each separately.

\subsection{Two Degrees-of-Freedom Hamiltonian Systems}

The geometry and dynamics of the phase space in a neighborhood of an index-one saddle was first worked out by Conley for two DoF systems in his studies of the restricted three body problem \cite{conley1969ultimate, conley1968low}, and we briefly describe that situation. The Lyapunov subcenter theorem \cite{moser1958generalization, kelley1967liapounov} states that for a range of energies above that of the index-one saddle (the precise range is not specified in the theorem)  there exists an unstable periodic orbit (UPO) having two dimensional stable and unstable manifolds. In the three dimensional energy surface the stable and unstable manifolds have the geometry of cylinders (``tubes’’) and these stable and unstable cylinders mediate transitions between the wells as we now describe.

As a result of the invariance of the stable and unstable manifolds, and their cylindrical structure, trajectories that start within the tubes must remain in the tubes for all positive and negative time. All trajectories inside the stable cylinder approach the UPO in positive time, pass through the regions bounded by the UPO, and exit the region through the unstable cylinder. The stable cylinder (and the unstable cylinder) has two ``branches’’ joined at the UPO. Considering a neighborhood of the UPO in the energy surface, one branch is in the reactant region and the other is in the product region.

All trajectories that react must lie in a branch of the stable cylinder. Trajectories starting in the branch of the stable cylinder lying in the reactant region correspond to forward reacting trajectories, i.e. trajectories that evolve from reactants to products in forward time. Trajectories starting in the branch of the stable cylinder lying in the product region correspond to backward reacting trajectories, i.e. trajectories that evolve from products to reactants in forward time. Hence the stable cylinders form the pathways for reaction in phase space. They can be used to monitor reacting trajectories in the following manner. For definiteness we consider the branch of the stable cylinder in the reactant region. Similar reasoning applies to the branch of the stable cylinder in the product region.

A dividing surface (DS) can be constructed to monitor, or ``count’’, the trajectories passing from reactants to products. In order for trajectories to be counted only once in their evolution from reactants to products they must cross the DS only once. A DS with this property is said to have the ``no-recrossing’’ property.  Seminal work of Pollak, Pechukas, and Child \cite{pollak1978,pechukas1977,pechukas1979, PollakChild80, PollakChildPechukas80} showed how to use this UPO to construct a DS with these properties, a so-called ``periodic orbit dividing surface’’, or PODS. The DS is topologically a 2 sphere. The equator of the 2 sphere is the UPO which divides that 2 sphere into two halves. One half is intersected by trajectories contained in the branch of the stable cylinder in the reactant region (forward reaction) and the other half is intersected by trajectories contained in the branch of the stable cylinder in the product region (backward reaction).

The two dimensional invariant cylinders can exhibit a complicated geometrical structure in the three dimensional energy surface. We describe a lower dimensional method for monitoring their evolution towards reaction as follows. In the region of the energy surface corresponding to reactants (a similar construction can be carried out for the product region)  we construct a two dimensional Poincar\'e section, i.e. a two dimensional surface where the Hamiltonian vector field is everywhere transverse to the surface. The Poincar\'e section is constructed in such a way that the stable cylinder intersects it in a topological circle \cite{DeVogelaere1955,PollakChild80}. The region bounded by this topological circle is referred to as a ``reactive island’’ \cite{marston1989}. The return map of the Poincar\'e section into itself is the map that associates to a point its first return to the Poincar\'e section under the flow generated by the Hamiltonian vector field. The inverse of this return map of the Poincar\'e section into itself is the map that associates to a point its first return to the Poincar\'e section under negative time, i.e. the point ``where it came from’’. 

If one considers the preimages of this reactive island by letting it evolve backwards under the inverse of the Poincar\'e map one obtains a reactive island on the Poincar\'e section that returns to the original reactive island, and then reacts by crossing the DS for positive time evolution. By repeating this construction we obtain a temporal ordered sequence of reactive islands which sequentially map to each other before reacting \cite{almeida1990}. We would like to remark that pathological intersections of the cylinders with the Poincar\'e section can occur, and are discussed in \cite{deleon1989}.

\begin{figure}[htbp]
    \centering
    \includegraphics[scale=0.4]{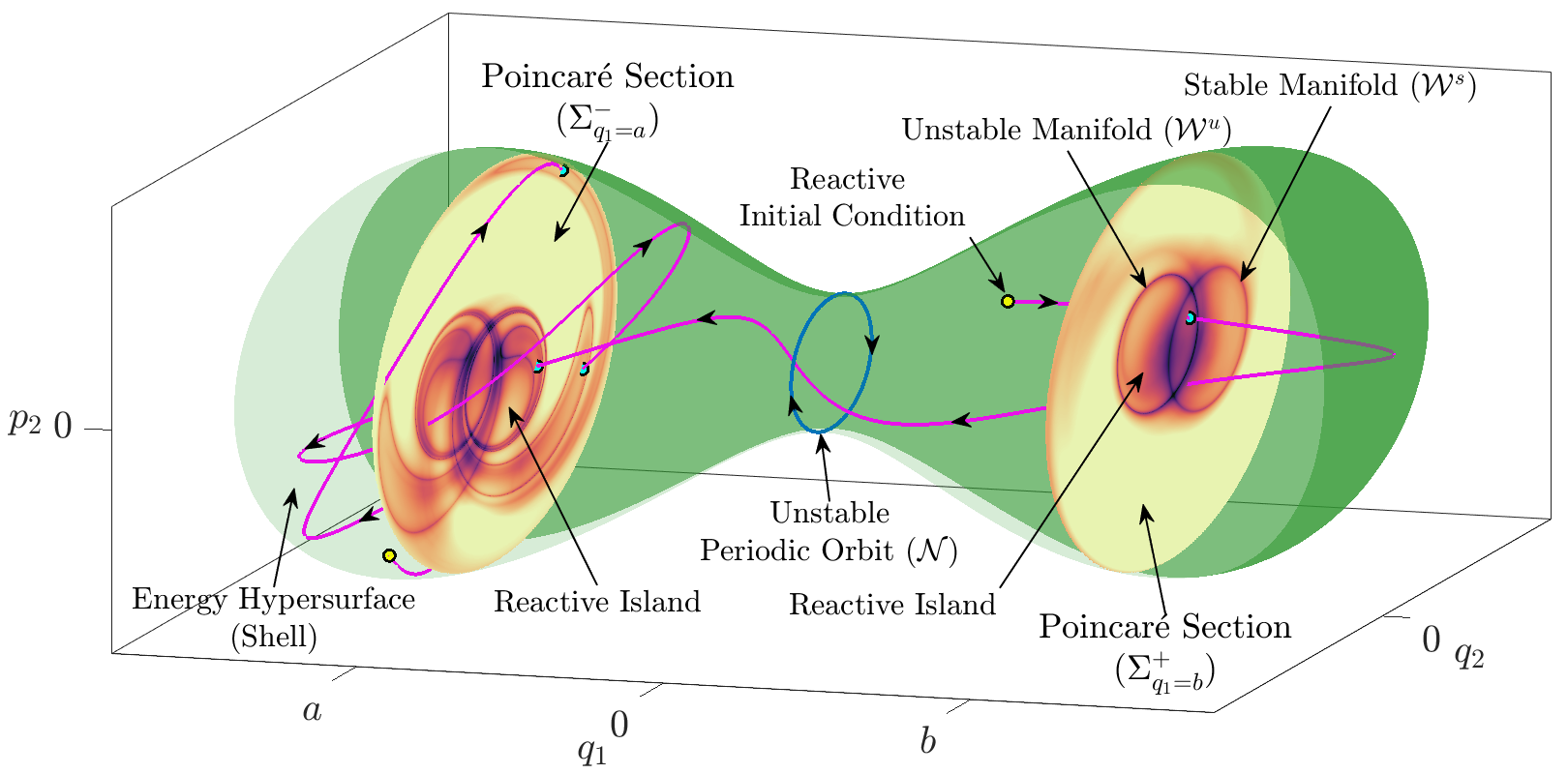}
    \caption{Reactive islands as revealed by the method of Lagrangian descriptors on the phase space of a coupled double well Hamiltonian system with 2 DoF. The cylindrical geometry of the stable and unstable manifolds emanating from the unstable periodic orbit in the bottleneck region is highlighted when LDs are applied on the two-dimensional Poincar\'e sections $\Sigma^{-}_{q_1 = a}$ and $\Sigma^{+}_{q_1 = b}$. A reactive trajectory is depicted in magenta and the intersections of its path with the phase space planes (taking into account the directionality of the sections) is marked with cyan dots. Arrows are included in the picture to indicate forward time evolution, and the green surface represents the projection of the boundary of the energy surface.}
    \label{fig:reactive_islands}
\end{figure}

We illustrate in Fig. \ref{fig:reactive_islands} the concept of reactive islands for 2 DoF Hamiltonian systems on a suitable Poincar\'e section. The intricate geometry of the reactive islands, obtained when the stable and unstable manifolds originating from an unstable periodic orbit intersect transversely an adequately chosen phase space section, is revealed using the method of Lagrangian descriptors calculated on that surface of section. This technique provides a clear and highly-detailed visualization of the reactive island picture, offering an understanding of the nonlinear interactions between the manifolds that exist in the phase space of the Hamiltonian system.

Consider for example a double well Hamiltonian defined by a potential energy surface, where the two wells are separated by an index-1 saddle that lies between them. In the diagram displayed in Fig. \ref{fig:reactive_islands}, we have represented the boundary of the energy surface where motion takes place in green, and $q_1$ is the reactive DoF of the system. We have depicted in magenta a reactive trajectory that evolves in forward time from the `reactant' to the `product' well regions of the PES, by moving across the phase space bottleneck that exists in the neighborhood of the UPO associated to the index-1 saddle. This trajectory starts inside the spherical cylinder corresponding to the stable manifold on the right side of the phase space, and once it crosses the dividing surface associated to the UPO, it enters the left well region following the unstable manifold. We show the reactive islands formed by the tube manifolds of the system by computing LDs on the Poincar\'e sections $\Sigma^{-}_{q_1 = a}$ and $\Sigma^{+}_{q_1 = b}$ taken on both well regions, where the plus and minus sign notation represents the directionality of the section, that is, $\dot{q}_1 > 0$ or $\dot{q}_1 < 0$ respectively. Moreover, on the section $\Sigma^{-}_{q_1 = a}$ LDs have been calculated using a longer integration time in order to unveil the successive sequence of reactive islands that arise when the manifolds intersect the section for several times. The deformation of the successive reactive islands observed in the Poincar\'e section is a fingerprint of the chaotic behavior of this system.

The flux of trajectories across the phase space bottleneck region, where the UPO associated to the index-1 saddle point is located, is a key ingredient for the computation of the classical reaction rate. The flux through the DS, $\varphi_{\text{DS}}$, can be calculated by means of applying Stokes theorem using the action integral \cite{mackay1990}, that is:
\begin{equation}
    \varphi_{\text{DS}} = \int_{\text{DS}} d\mathbf{p} \wedge d\mathbf{q} = \oint_{\text{UPO}} \mathbf{p} \cdot d\mathbf{q}
\end{equation}
where $\mathbf{q}$ represents the configuration space variables and $\mathbf{p}$ the canonically conjugate momenta. Following the Poincar\'e-Cartan theorem \cite{arnold1978}, we have:
\begin{equation}
    \oint_{\text{UPO}} \mathbf{p} \cdot d\mathbf{q} = \oint_{\Gamma} \mathbf{p} \cdot d\mathbf{q},
\end{equation}
for any curve $\Gamma$ that results from a transverse intersection of the stable or unstable manifolds of the UPO with a Poincar\'e section. Hence, the flux across the dividing surface in the phase space bottleneck is the same as the flux across any of the reactive islands.

A theory of reaction dynamics for two DoF systems based on the geometry of these stable and unstable cylinders and ``reactive islands'' \cite{marston1989} was developed in the late 1980’s and 90’s in \cite{deleon1989,marston1989,almeida1990, deleon91a, deleon91b, deleon92, mehta1992classical, fair1995}, although as remarked in \cite{almeida1990}, the ideas of fluxes of trajectories in tubes in phase space and their circular cross sections were considered in \cite{DeVogelaere1955, PollakChild80}.  The phase space structures and geometrical considerations that we have described above are very specific to two DoF Hamiltonian systems \cite{almeida1990}. However, in essentially all of the papers noted above the need and interest for an analogous theory is described as ``future work’’. However, such future work has not appeared. There are two reasons for this. One is that new types of dynamical objects are required, and the other is that a numerically efficient method for generating these structures is required. In recent years progress has been made along both lines that enable us to realize a reactive island theory for three DoF systems, which we now describe.

\subsection{Three Degrees-of-Freedom Hamiltonian Systems}

Three DoF Hamiltonian systems have a 6-dimensional phase space and dynamics is constrained to a 5-dimensional energy surface. An UPO has 3-dimensional stable and unstable manifolds in the 5-dimensional energy surface.  Hence, they are not codimension one, i.e. one less dimension than the energy surface, and therefore they cannot trap trajectories or act as barriers in the phase space. 

It has been shown that an appropriate generalization of an UPO for 3 DoF Hamiltonian systems is a normally hyperbolic invariant 3-sphere having 4-dimensional stable and unstable manifolds in the 5-dimensional energy surface \cite{wiggins90, Komatsuzaki97, Komatsuzaki00, wiggins2001, uzer2002,Komatsuzaki02a,wig2016}.  The stable and unstable manifolds have the geometry of a spherical cylinder, or ``spherinder’’, see e.g. https://en.wikipedia.org/wiki/Spherinder. The 3-sphere can be used to construct a DS with the same properties as the PODS for 2 DoF Hamiltonian systems that we described earlier. 

However, the main issue is how would this approach be implemented in specific systems. In particular, how can one determine the existence and compute the normally hyperbolic invariant 3-sphere, compute its 4-dimensional stable and unstable manifolds (the ``spherinders’’), and determine, and understand, the geometry of their intersection with a 4-dimensional Poincar\'e section? These intersection sets are 3-dimensional on the Poincar\'e section and are the analog of the reactive islands for the 2 DoF Hamiltonian systems discussed earlier.

For 2 DoF Hamiltonian systems there is a rich body of knowledge concerning computing UPOs and their stable and unstable manifolds.  However, the generalization of these 2 DoF techniques to the 3 DoF setup described above is simply not practical. For example, suppose we had a realization of the normally hyperbolic invariant 3-sphere adequate for computational purposes (this is a big step that we are just assuming, for the moment).  The next step would be to compute its 4-dimensional stable and unstable manifolds. The analog of the situation for 2 DoF would be to generate a 4-dimensional mesh of initial conditions near the linearized unstable directions of the sphere and propagate them in time in the 5-dimensional energy surface. However, as it is propagated gaps would quickly develop in the mesh and repeated remeshing in space after intervals of time would be required. This has been carried out for the computation of 2-dimensional stable and unstable manifolds in 3 dimensions, but not in higher dimensions.

We note that Poincar\'e maps  ``work’’ for 2 DoF systems because they are 2-dimensional and it is relatively easy to compute many trajectories that return to the Poincar\'e section and give rise to structures that can be related to dynamical concepts. For 3 DoF systems the Poincar\'e section is 4-dimensional and if sufficient trajectories returned it is not obvious how one could extract ``recognizable’’ structures in this 4-dimensional space. Of course, one could attempt to extract structures in the 4-dimensional Poincar\'e section by considering 2-dimensional slices of the Poincar\'e section. However, the probability that a trajectory with initial conditions on this 2-dimensional slice returning to this slice in 5 dimensions is zero. For completeness we add that 2-tori around stable periodic orbits were investigated on 4-dimensional Poincar\'e sections in 3 DoF systems using color as the fourth dimension \cite{Patsis1994} and on ``thickened'' 3-dimensional slices in 4-dimensional maps \cite{lange2014}. It is not clear how these approaches could be used for 3-dimensional objects.

The relatively new trajectory diagnostic known as ``Lagrangian descriptors’’ is giving very promising results for uncovering phase space structures in high-dimensional systems. The method was originally developed in the context of Lagrangian transport studies in fluid dynamics \cite{madrid2009ld, mancho2013lagrangian,lopesino2017}, but the wide applicability of the method has recently been recognized in chemistry \cite{craven2015lagrangian,craven2016deconstructing,craven2017lagrangian,junginger2017chemical,feldmaier2017obtaining,revuelta2019unveiling,krajnak2020manifld,montoya2020revealing,montoya2020phase}. The method is straightforward to implement computationally \cite{ldbook2020}, the interpretation is clear, and it provides a “high resolution” method for exploring high-dimensional phase space with low-dimensional slices \cite{naik2019a, naik2019b}.  

Phase space structures consist of trajectories. But as discussed above, for high-dimensional phase space trajectory approaches for revealing phase space structure are problematic and prone to issues of interpretation since a tightly grouped set of initial conditions may result in trajectories that become “lost” with respect to each other in phase space. The method of Lagrangian descriptors turns this problem on its head by emphasizing the initial conditions of trajectories, rather than the precise location of their futures and pasts, after a specified amount of time. In fact, Lagrangian descriptors have already been used to reveal reactive islands in 2 DoF Hamiltonian systems \cite{patra2015, patra2018, naik2020, krajnak2020manifld}. A low-dimensional “slice” of phase space can be selected and sampled with a grid of initial conditions of high resolution. Since the phase space structure is encoded in the initial conditions of the trajectories, no resolution is lost as the trajectories evolve in time. In this paper we will show how this can be done in order to realize the setting of reactive islands for 3 DoF Hamiltonian systems. We will show how the reactive islands can be extracted from Lagrangian descriptor values and we will be able to calculate flux by means of quantifying the volume enclosed by them.

Recall that, as we have discussed above, reactive islands in 3 DoF systems have the topology of $S^3$. To facilitate the work we will decompose this high-dimensional structures into a family of two-dimensional sphere-like objects, $S^2$, embedded in 3-dimensional slices, that we triangulate and accumulate their contributions into the flux integral. Furthermore, we will show how to calculate the volume enclosed by the intersection of two reactive islands, a crucial step towards determining reaction rates.

Detecting these structures in this high-dimensional setup, and using them to compute flux, is a challenging task that we address in this work. We provide a procedure for systematically carrying out these computations, and illustrate how this algorithm can be applied to a simple benchmark Hamiltonian system with 3 DoF, a double well potential which describes a reactive DoF quadratically coupled to two harmonic oscillators. 

This paper is outlined as follows. Section \ref{sec2} introduces the 3 DoF double well system that we will use throughout this work as a simple model to study the concept of reactive islands, and also for analyzing the computation of flux. In Sec. \ref{sec3} this Hamiltonian model is extended further by adding a quadratic coupling between the reactive and bath (i.e. harmonic oscillator) degrees-of-freedom. Section \ref{sec4} focuses on describing how to select an adequate Poincar\'e surface of section so that the reactive island approach for calculating flux works. Section \ref{sec5} is devoted to explaining how the method of Lagrangian descriptors can be used to extract the high-dimensional reactive islands (and hence visualize their geometry) formed by the intersections of the stable and unstable manifolds in the system. This step is crucial in order to develop an algorithm that allows us to determine fluxes for the uncoupled and coupled system scenarios from the reactive islands detected by means of Lagrangian descriptors. The computation of flux for the uncoupled Hamiltonian is described in Sec. \ref{sec6}, whereas Sec. \ref{sec7} does the same for the coupled system. Finally, in Sec. \ref{conc} we present the conclusions of this work and discuss some ideas for future research lines where the methodology we have developed for identifying high-dimensional reactive islands using Lagrangian descriptors could provide relevant insights and lead to interesting results. 


\section{Double Well Potential Hamiltonian with Three Degrees-of-Freedom}
\label{sec2}

In this section we present a benchmark 3 DoF example that allows analytical calculation of the phase space structures that play the central role for a reactive island theory in 3 DoF Hamiltonian systems. It is an uncoupled 3 DoF Hamiltonian system consisting on a 1 DoF double well PES and two harmonic oscillators. This is the natural generalization of the motivational example in 
\cite{almeida1990} and follows the spirit of \cite{wig2016}. In particular, we consider the following Hamiltonian system:
\begin{equation}
\mathcal{H}(q_1,q_2,q_3,p_1,p_2,p_3) = \dfrac{p_1^2}{2} + \dfrac{q_1^4}{4} - \dfrac{q_1^2}{2} + \dfrac{\omega_2}{2} \left(q_2^2+p_2^2\right) + \dfrac{\omega_3}{2} \left(q_3^2+p_3^2\right) \;,
\end{equation}
where $\omega_2,\,\omega_3 > 0$ are the angular frequencies of the harmonic oscillators. Notice that we have written the contributions of the harmonic oscillators to the Hamiltonian differently from the classical expression containing the squares of the frequencies. In fact, both expressions are equivalent by a canonical transformation. This approach is more convenient to use and is similar to the convention followed in chemistry when working with mass-weighted coordinates \cite{wilson1980}. The corresponding Hamilton's equations of motion that describe the dynamics of the system are as follows:
\begin{equation}
\begin{cases}
\dot{q}_1 = \dfrac{\partial \mathcal{H}}{\partial p_1} = p_1 \\[.4cm]
\dot{q}_2 = \dfrac{\partial \mathcal{H}}{\partial p_2} = \omega_2 p_2 \\[.4cm]
\dot{q}_3 = \dfrac{\partial \mathcal{H}}{\partial p_3} = \omega_3 p_3 \\[.4cm]
\dot{p}_1 = - \dfrac{\partial \mathcal{H}}{\partial q_1} = q_1 - q_1^3 \\[.4cm]
\dot{p}_2 = - \dfrac{\partial \mathcal{H}}{\partial q_2} = -\omega_2 q_2 \\[.4cm]
\dot{p}_3 = - \dfrac{\partial \mathcal{H}}{\partial q_3} = -\omega_3 q_3
\end{cases}
\end{equation}

We will construct reactive islands for a fixed energy. Hence, the 5-dimensional energy surface is given by:
\begin{equation}
\mathcal{E}_{E} = \left\{(q_1,q_2,q_3,p_1,p_2,p_3) \in  \mathbb{R}^6 \; \Big| \;  \dfrac{p_1^2}{2} + \dfrac{q_1^4}{4} - \dfrac{q_1^2}{2} + \dfrac{\omega_2}{2} \left(q_2^2 + p_2^2\right) + \dfrac{\omega_3}{2} \left(q_3^2 + p_3^2\right) = E \right\} \,.
\end{equation}
The following results are proven in \cite{wig2016}.
The isoenergetic normally hyperbolic invariant manifold (NHIM) is given by:
\begin{equation}
\mathcal{N}_{E} = \left\{(q_1,q_2,q_3,p_1,p_2,p_3) \in  \mathbb{R}^6 \; \Big| \;  \dfrac{\omega_2}{2} \left(q_2^2 + p_2^2\right) + \dfrac{\omega_3}{2} \left(q_3^2 + p_3^2\right) = E \, , \,  q_1 = p_1 = 0 \right\} \,.
\end{equation}
and it has the form of a 3-sphere in the 5-dimensional energy surface. In this example the stable and unstable manifolds  (``spherinders’’) of the NHIM are coincident and are given by:
\begin{equation}
\mathcal{W}^{s,u}\left(\mathcal{N}_{E}\right) = \left\{ (q_1, q_2, q_3, p_1, p_2, p_3) \in \mathbb{R}^6 \; \Big| \;  \dfrac{\omega_2}{2} (q_2^2 + p_2^2) + \dfrac{\omega_3}{2} (q_3^2 + p_3^2) = E \; , \; \dfrac{p_1^2}{2} + \dfrac{q_1^4}{4} - \frac{q_1^2}{2} = 0 \; , \; q_1 \neq 0 \right\} \,.
\end{equation}
They are 4-dimensional in the 5-dimensional energy surface. The dividing surface ($q_1 = 0$) is given by:
\begin{equation}
\mathcal{D}_{E} = \left\{ (q_1, q_2, q_3, p_1, p_2, p_3) \in \mathbb{R}^6 \; \Big| \; \dfrac{1}{2} p_1^2 + \dfrac{\omega_2}{2} \left(q_2^2 + p_2^2\right) + \dfrac{\omega_3}{2} \left(q_3^2 + p_3^2\right) = E \; , \; q_1 = 0 \right\} \,.
\end{equation}
and has the form of a 4-sphere in the 5-dimensional energy surface. Its equator is given by $p_1 = 0$, which is the NHIM.


\subsection{Intersection of \texorpdfstring{$\mathcal{W}^{s,u}\left(\mathcal{N}_{E}\right)$}{TEXT} with a hyperplane}

\begin{figure}[htbp]
    \centering
    \includegraphics[width=0.35\textwidth]{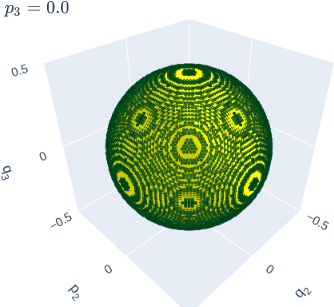} \hfil
    \includegraphics[width=0.35\textwidth]{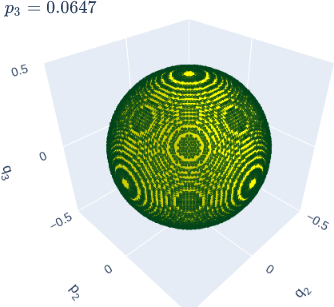} \\
    \includegraphics[width=0.35\textwidth]{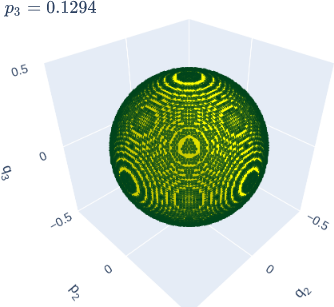} \hfil
    \includegraphics[width=0.35\textwidth]{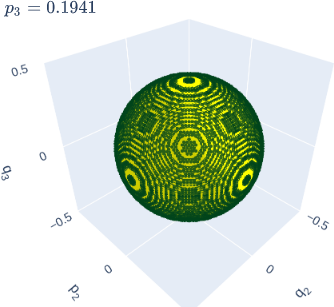} \\
    \includegraphics[width=0.35\textwidth]{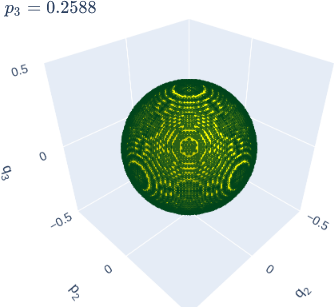} \hfil
    \includegraphics[width=0.35\textwidth]{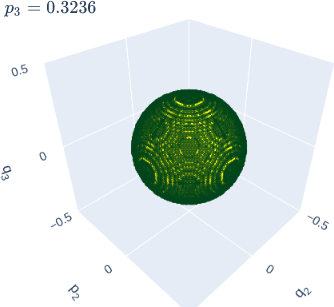} 
    \caption{Section of the stable invariant manifold $\Sigma_{0.5} \cap\mathcal{W}^s$ of the uncoupled system with $E=0.1$ and $\omega_2=\omega_3=1$. The section consists of several $p_3=\text{const}>0$ slices ($p_3<0$ is related by symmetry) of which the first six are shown. The analytically calculated manifold in yellow is overlayed with green points calculated numerically as detailed in Sec. \ref{sec5}.}
    \label{fig:DW_spheres}
\end{figure}

As discussed in the introduction, reactive islands are the intersection of the stable manifold of the normally hyperbolic invariant 3-sphere (``spherinder’’)  with a 4-dimensional Poincar\'e section in the energy surface. In particular, we will show that in an energy surface, a transverse hyperplane intersects $\mathcal{W}^{s,u}\left(\mathcal{N}_{E}\right)$ in a 3-sphere. We define a hyperplane that intersects $\mathcal{W}^{s,u}\left(\mathcal{N}_{E}\right)$ transversely by: 
\begin{equation}
\Sigma_{0.5} = \left\{ (q_1, q_2, q_3, p_1, p_2, p_3) \in \mathbb{R}^6 \; \Big| \; \mathcal{H}(q_1,q_2,q_3,p_1,p_2,p_3) = E \,,\, q_1 = -0.5 \, , \,\dot{q}_1 > 0 \right\} \,.
\end{equation}
then we have:
\begin{equation}
\Sigma_{0.5} \cap \mathcal{W}^{s,u}\left(\mathcal{N}_{E}\right) 
= \left\{ (q_1, q_2, q_3, p_1, p_2, p_3) \in \mathbb{R}^6 \; \Big| \;  \dfrac{\omega_2}{2} (q_2^2 + p_2^2) + \dfrac{\omega_3}{2} (q_3^2 + p_3^2) = E \; , \; q_1 = -0.5 \; , \; p_1 = \sqrt{\dfrac{7}{32}} \right\}.
\label{eq:Sigma_cap_manif}
\end{equation}
which is a 3-sphere, the higher-dimensional analogue of the reactive islands shown on $\Sigma^{-}_{q_1 = a}$ and $\Sigma^{+}_{q_1 = b}$ in Fig. \ref{fig:reactive_islands}. In Fig. \ref{fig:DW_spheres} we show several $p_3=\text{const}>0$ slices of this 3-sphere.

\section{The Double Well Potential with Quadratic Coupling}
\label{sec3}

We now introduce a quadratic coupling term between the reactive DoF, $q_1$, and each of the bath DoF, $q_2$ and $q_3$, for the model system described in Sec. \ref{sec2}. The coupled Hamiltonian has the form:
\begin{equation}
\mathcal{H}(q_1,q_2,q_3,p_1,p_2,p_3) = \dfrac{p_1^2}{2} + \dfrac{q_1^4}{4} - \dfrac{q_1^2}{2} + \dfrac{\omega_2}{2} \left(q_2^2+p_2^2\right) + \dfrac{\omega_3}{2} \left(q_3^2+p_3^2\right) + \dfrac{\varepsilon}{2} \left(q_1 - q_2\right)^2 + \dfrac{\varepsilon}{2} \left(q_1 - q_3\right)^2\;,
\end{equation}
where the degrees of freedom are coupled quadratically with the coupling parameter $\varepsilon > 0$. This system gives rise to the following Hamilton's equations:
\begin{equation}
\begin{cases}
    \dot{q}_1 = \dfrac{\partial \mathcal{H}}{\partial p_1} = p_1, \\[.4cm]
    \dot{q}_2 = \dfrac{\partial \mathcal{H}}{\partial p_2} = \omega_2 p_2 \\[.4cm]
    \dot{q}_3 = \dfrac{\partial \mathcal{H}}{\partial p_3} = \omega_3 p_3 \\[.4cm]
    \dot{p}_1 = - \dfrac{\partial \mathcal{H}}{\partial q_1} = q_1 - q_1^3 + \varepsilon \left(q_2 - q_1\right) + \varepsilon \left(q_3-q_1\right) \\[.4cm]
    \dot{p}_2 = - \dfrac{\partial \mathcal{H}}{\partial q_2} = -\omega_2 q_2 + \varepsilon \left(q_1-q_2\right) \\[.4cm]
    \dot{p}_3 = - \dfrac{\partial \mathcal{H}}{\partial q_3} = -\omega_3 q_3 + \varepsilon \left(q_1-q_3\right)
\end{cases}
\;.
\end{equation}

While the NHIM $\mathcal{N}$ associated with the index-1 saddle is still a $3$-sphere, its location and shape depends on the coupling strength $\varepsilon$. It can be approximated by a nearly invariant linear combination of $q_1,q_2,q_3$ on the energy surface $\mathcal{H}=E$. For the sake of simplicity we derive this linear combination for $\omega_2=\omega_3=1$ as
\begin{equation}
 cq_1-\varepsilon(q_2+q_3)=0 \,,
\end{equation}
where
\begin{equation}
c = \dfrac{1}{2}\left(\varepsilon-2-\sqrt{\left(\varepsilon-2\right)^2+8\varepsilon^2}\right) ,
\label{eq:c}
\end{equation}
follows from the near-invariance under the equations of motion linearized around $q_1=0$ (or the origin):
\begin{equation}
c  p_1-\varepsilon(p_2+p_3) = 0 \,,\quad
c  \dot{p}_1-\varepsilon(\dot{p}_2+\dot{p}_3) = 0.
\end{equation}
We define the dividing surface constructed using the approximated NHIM as:
\begin{equation}
\mathcal{D}_{E} = \left\{ (q_1, q_2, q_3, p_1, p_2, p_3) \in \mathbb{R}^6 \; \Big| \; \mathcal{H}(q_1,q_2,q_3,p_1,p_2,p_3) = E \; , \; cq_1-\varepsilon(q_2+q_3)=0 \right\} \,.
\label{eq:CDW_DS}
\end{equation}
In all of the following calculations for the coupled system we consider $\varepsilon=0.1$ and $\omega_2=\omega_3=1$.


\section{Choice of surface of section and tangencies}
\label{sec4}

For the reactive island approach to work correctly, it is crucial that the hyperplane $\Sigma$ is defined transversely to the investigated invariant manifold. It is well known that discontinuities occur when a trajectory on an invariant manifold is tangent to the hyperplane in systems with $2$ DoF \cite{almeida1990}, and the only difference with 3 DoF systems involves the dimensionality of the tangent set, which can vary from $0$ to $2$.

\begin{figure}[htbp]
  \includegraphics[width=0.3\textwidth]{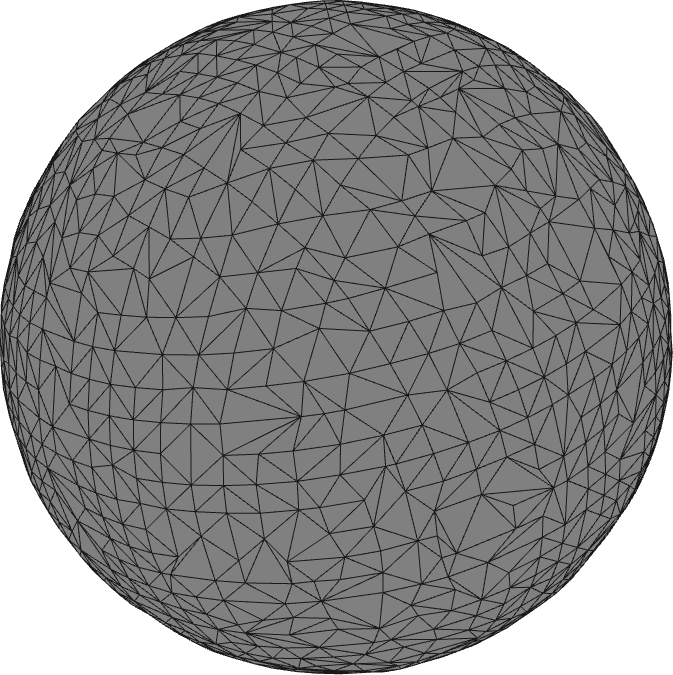}\hfil
  \includegraphics[width=0.3\textwidth]{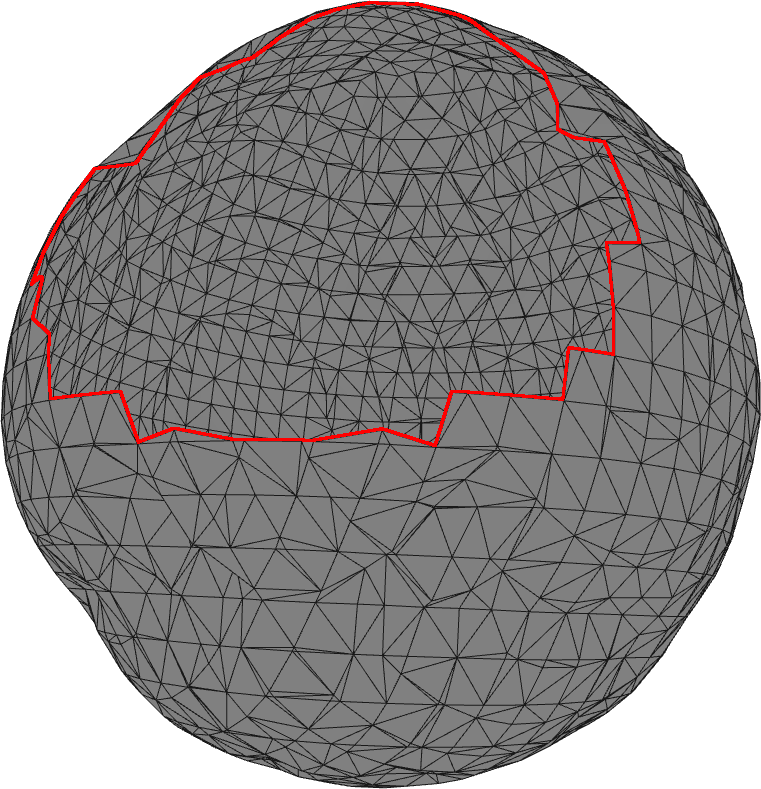}
  \caption{Transverse intersection of $\mathcal{W}^{s}$ (left) and non-transverse intersection of $\mathcal{W}^{u}$ (right) with $\Sigma_{0.5}$ and $\left\{p_3 \approx 1/3\right\}$ in the coupled system with $\varepsilon=0.1$. The hole in the non-transverse intersection is highlighted in red.}
  \label{fig:nontransverse}
\end{figure}

While a transverse intersection of $\Sigma$ and $\mathcal{W}^{s,u}$ is a 3-sphere, see Eq. \eqref{eq:Sigma_cap_manif}, a non-transverse intersection causes this sphere to have holes as we show in Fig. \ref{fig:nontransverse}. As a consequence, a portion of the invariant manifold and the energy surface volume it encloses do not reach $\Sigma$, which leads to errors of qualitative and quantitative nature. We remark that one can always estimate the likelihood of tangent intersections by considering the Hamiltonian vector field on the 3-sphere, for example how close $p_1$ is to $0$ in case of $\Sigma$.


\section{Approximating Reactive Islands With Lagrangian Descriptors}
\label{sec5}

To approximate invariant manifolds, we employ the method of Lagrangian descriptors (LDs). This technique has its origins in fluid dynamics, more precisely in Lagrangian transport phenomena \cite{madrid2009ld}, and has since become a popular tool for exploring the phase space of dynamical systems. For a detailed exposition on the method, with examples on how it can be used to reveal the phase space of Hamiltonian systems, refer to \cite{ldbook2020}.

A Lagrangian descriptor is a non-negative function $f$ of the phase space variables that is integrated along trajectories of a system for a fixed amount of time $\tau$ forward and backward, assigning every initial condition $(\mathbf{q}_0,\mathbf{p}_0)$ at time $t = t_0$ a value given by:
\begin{equation}
 L(\mathbf{q}_0,\mathbf{p}_0,t_0,\tau) = \int^{t_0+\tau}_{t_0 - \tau} f(\mathbf{q},\mathbf{p};\mathbf{q}_0,\mathbf{p}_0) \, dt \, .
\end{equation}
Among the common choices for the function $f$ are arclength, $L^p$ norm ($p\geq1$) or semi-norm ($p<1$) and action \cite{madrid2009ld,mancho2013lagrangian,lopesino2017}. If $f$ is chosen appropriately, initial conditions leading to qualitatively different dynamics, for example oscillation and rotation, attain different ranges of values for the Lagrangian descriptor. Abrupt transitions in the values of $L$ correspond to (codimension-$1$) invariant manifolds that form the boundary of areas of qualitatively different dynamics.
 
Here we follow \cite{krajnak2020manifld} to directly approximate specific invariant manifolds using some knowledge of NHIMs in the systems. We are interested in invariant manifolds of codimension-$1$ in the energy surface $\mathcal{H}=E$, that are asymptotic to $\mathcal{N}_{E}$, which is of codimension-$2$ in the energy surface and codimension-$3$ in phase space. Any codimension-$3$ object can be written as the intersection of three hypersurfaces, namely the energy surface and two other distinct hypersurfaces $f_1(q,p)=0$ and $f_2(q,p)=0$. For example, in the uncoupled system in Sec. \ref{sec2}, these can be $q_1 = 0$ and $p_1 = 0$. Due to the invariance of $\mathcal{N}_{E}$, for any initial condition $(q,p)$ on $\mathcal{N}_{E}$ we have that:
\begin{equation}
\int^{\pm\infty}_{0} |\dot{f_i}(q,p)| \, dt = 0 \;,\quad i = 1,2 \;.
\label{eq:LDinfty}
\end{equation}

For initial conditions on the stable invariant manifold $\mathcal{W}^{s}(\mathcal{N}_{E})$ (unstable invariant manifold $\mathcal{W}^{u}(\mathcal{N}_{E})$) $f_i\rightarrow 0$ as $t\rightarrow \infty$ ($t\rightarrow-\infty$) and the integral in Eq. \eqref{eq:LDinfty} does not vanish, but attains a local minimum. Provided the integration time $\tau$ is sufficiently large, a local minimum is attained for finite values. For having the most accurate result, it is important to select initial conditions in such a way, that trajectories on the manifolds converge to $\mathcal{N}_{E}$ without any preceding complicated evolution. In the system with $2$ DoF considered in \cite{krajnak2020manifld}, accurate results were obtained this way when $\tau$ was similar to the period of the periodic orbits. We remark that it suffices to integrate with lower precision than that used to approximate NHIMs in the system.
 
To find points on the invariant manifolds for the uncoupled double well system, see Fig. \ref{fig:DW_spheres}, we uniformly sample the surface $\Sigma_{0.5}$. Since $\mathcal{W}^{s}(\mathcal{N}_{E})$ and $\mathcal{W}^{u}(\mathcal{N}_{E})$ coincide, we only need to approximate one of them. As argued above, it is important that trajectories immediately approach $\mathcal{N}_{E}$, we therefore calculate Lagrangian descriptor values for the stable manifold.

For the coupled double well with $\varepsilon = 0.1$ considered here, $\mathcal{W}^{u}$ does not intersect $\Sigma_{0.5}$ transversely, see Sec. \ref{sec4} and Fig. \ref{fig:nontransverse}. For simplicity, in the coupled system we always use the hyperplane
\begin{equation}
\Sigma_{0.7} = \left\lbrace (q_1,q_2,q_3,p_1,p_2,p_3) \in \mathbb{R}^6 \; \Big| \; \mathcal{H}(q_1,q_2,q_3,p_1,p_2,p_3) = E \,,\, q_1 = -0.7 \,,\, \dot{q}_1 > 0\right\rbrace
\end{equation}
In general, a surface constructed using the stable object associated with the local minimum at $q_1=\pm\sqrt{1-2\varepsilon}$ will intersect $\mathcal{W}^{s,u}$ transversely. From $\Sigma_{0.7}$, we integrate trajectories until $q_1 = -0.5,\, \dot{q}_1>0$ in forward time and $q_1=-0.5,\, \dot{q}_1<0$ in backward time and subsequently we calculate Lagrangian descriptor values for a fixed time interval starting at $q_1=-0.5$. This way we omit the contribution of the dynamics near the local minimum from Lagrangian descriptor values, making them comparable to values obtained in the uncoupled system. We remark that higher order reactive islands can be calculated analogously using trajectories that return to $\Sigma_{0.7}$ the corresponding number of times.
 
Points closer to the invariant manifolds attain lower LD values. In the uncoupled system, it is possible to explicitly calculate the values of LDs along invariant manifolds, we then only considered points whose LD did not exceed, for example, twice the actual value. In general a suitable cut-off value can be determined using the lowest attained LD value or from a histogram of calculated LD values.

The resulting cloud of points approximates the intersection of the invariant manifold with $\Sigma_{0.5}$ or $\Sigma_{0.7}$, a $3$-sphere, and we perform calculations on its triangulation. For a more understandable presentation, we cut the sphere in (disjoint) slices of the form $p_3=\text{const}$, resulting in a continuum of $2$-spheres embedded in $3$-dimensional space. We triangulate each $2$-sphere separately using the function \verb|geometry.TriangleMesh.create_from_point_cloud_poisson| of \verb|open3d| \cite{Zhou2018}, which reconstructs a smooth surface approximated by oriented triangles as shown in Fig. \ref{fig:nontransverse}. This Poisson reconstruction implementation creates a convex hull, with the parameter \emph{depth} balancing smoothness (low values) and level of detail (high value). Throughout this work we use $depth=5$, where the default is value $8$. The resulting mesh may be simplified to an arbitrary number of triangles. We verify the connectedness of the resulting triangulated surface using the built in method of mesh objects \verb|is_watertight| and its Euler characteristic ($2$ for a sphere) using \verb|euler_poincare_characteristic|. Software for triangulation of higher-dimensional manifolds exists or is in development and should be used for treatment of higher-dimensional systems.


\section{Calculating flux}
\label{sec6}

If we select the Poincar\'e section:
\begin{equation}
\Sigma = \left\lbrace (q_1,q_2,q_3,p_1,p_2,p_3) \in \mathbb{R}^6 \; \Big| \; \mathcal{H}(q_1,q_2,q_3,p_1,p_2,p_3) = E \, , \, q_1 = \text{const} \, , \, \dot{q}_1 > 0\right\rbrace \,,
\end{equation}
where $p_1$ is implicitly given by energy, the volume of the (unidirectional) flux of energy surface volume enclosed by $\mathcal{W}^{s}$ across a section $\Sigma$ is defined by
 \begin{equation}
  \int_{\mathcal{W}^{s}\cap\Sigma} p_2 \; dq_2 \wedge dp_3 \wedge dq_3,
  \label{eq:flux}
 \end{equation}
 and analogously for $\mathcal{W}^{u}$. We remark that his flux is a discrete time flux associated with the return map, that is a volume per iteration of a return map, which is different to the volume per unit of time $t$.
 
 When cut in finitely many $p_3=\text{const}$ slices, integral \eqref{eq:flux} is approximated by
 \begin{equation}
  \sum_i (p_{3_i}-p_{3_{i-1}}) \, \int_{S_i} p_2 \, dq_2 \wedge dq_3 \, ,
 \end{equation}
 where $S_i=\mathcal{W}^{s}\cap\Sigma\cap\{p_3=p_{3_i}\}$ and upon triangulation of $S_i$ this becomes
 \begin{equation}
  \sum_i (p_{3_i}-p_{3_{i-1}}) \sum\limits_\Delta \int\limits_\Delta p_2 \, dq_2 \wedge dq_3.
 \end{equation}
 Given the planarity of individual triangles and adjustable triangulation density, approximating $p_2$ on the triangle by a constant value $\bar{p}_2$, we can further simplify the integral
  \begin{equation}
  \int\limits_\Delta p_2 \, dq_2 \wedge dq_3 \approx \bar{p}_2\int\limits_\Delta dq_2 \wedge dq_3,
 \end{equation}
 where $\int\limits_\Delta dq_2 \wedge dq_3$ is the area of the projection of the triangle onto the $q_2,q_3$-plane.
  
 For the uncoupled double well, the integral in Eq. \eqref{eq:flux} is equal to
 \begin{equation}
    \int_{\mathcal{N}_{E}} dp_2 \, \wedge dq_2 \wedge dp_3 \wedge dq_3 = \text{vol}\left(\mathcal{N}_{E}\right) = \dfrac{2\pi^2E^2}{\omega_2 \omega_3},
    \label{eq:uncoupled_flux_nhim}
 \end{equation}
 which for $E=0.1$ and $\lambda=\omega_2=\omega_3=1$ is approximately $0.1974$. Table \ref{tab:uncoupled} summarises the flux across the first reactive island estimated by our algorithm for a $60\times60\times60\times10$ uniform grid in $\{0 \leq q_2, p_2,q_3,p_3 \leq c_{0.5}\}$  (negative values are obtained via symmetry) on $\Sigma_{0.5}$, where $c_{0.5} = \sqrt{2E+7/32}$ is the maximal value attainable by any of the harmonic oscillators for $q_1 = \pm 0.5$. Following the Poincar\'e-Cartan theorem \cite{arnold1978} the flux across the reactive island is the same as \eqref{eq:uncoupled_flux_nhim}.
 We used the Lagrangian descriptor
 \begin{equation}
 \int^{\tau}_{0} |\dot{q}_1| \, dt \,,
  \label{eq:LD6}
 \end{equation}
 where $\tau$ is chosen close to the period $2\pi$ of orbits on the NHIM.

\begin{table}[h]
   \begin{tabular}{l l|c c c c c}
   \multicolumn{2}{c}{} & \multicolumn{5}{c}{cut-off}\\
    & & 0.55 & 0.6 & 0.8 & 1.0 & 1.5\\
   \hline
   \multirow{3}{*}{\rotatebox{90}{$\tau$}}
    & 5 & 0.1982 & 0.1983 & 0.1983 & 0.2006 & 0.2373\\
    & 6 & 0.1976 & 0.1976 & 0.1977 & 0.1977 & 0.1989\\
    & 7 & - & - & 0.1972 & 0.1972 & 0.1972\\
    & 8 & - & - & 0.1960 & 0.1971 & 0.1972\\
\end{tabular}
\caption{\label{tab:uncoupled}Estimated flux across the first reactive island in the uncoupled system using LD in Eq. \eqref{eq:LD6} for selected $\tau$ and cut-off. ``-'' stands for unreliable estimate due to holes in the triangulated surface and deviating Euler characteristics.}
\end{table}


\section{Flux in the coupled double well}
\label{sec7}

\begin{figure}[htbp]
  \begin{center}
   \includegraphics[width=0.45\textwidth]{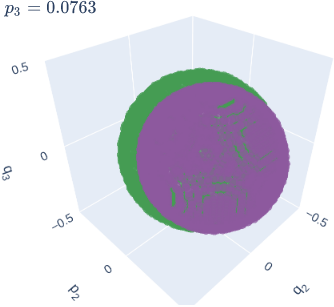}
  \end{center}
  \caption{The stable $\mathcal{W}^{s}$ (green) and unstable $\mathcal{W}^{u}$ (purple) manifolds in the coupled system no longer coincide.}
  \label{fig:CDW_spheres}
\end{figure}

In the uncoupled system, $\mathcal{W}^{s}$ and $\mathcal{W}^{u}$ coincide, leading all enclosed trajectories back and forth between the wells. In the coupled system, $\mathcal{W}^{s}$ and $\mathcal{W}^{u}$ do not coincide (see Fig. \ref{fig:CDW_spheres}). The analogous structure for systems with 2 DoF has been described in \cite{MacKay84} as a partial barrier, which leads some trajectories from one well to the other, while trapping other trajectories in the wells.
 
Figure \ref{fig:CDW_spheres} shows a single $p_3=\text{const}$ slice of the first forward and backward reactive islands on $\Sigma_{0.7}$. Recall that all trajectories inside the first forward reactive island (enclosed by $\mathcal{W}^{s}$) are carried across the dividing surface $\mathcal{D}_{E}$ defined in Eq. \eqref{eq:CDW_DS} in forward time without returning to $\Sigma_{0.7}$. This includes trajectories on $\mathcal{W}^{u}$. Following \cite{krajnak2020hydrogen}, we can determine which part of the first forward reactive island lies inside the first forward reactive island by integrating trajectories on $\mathcal{W}^{u}$ forward in time - ones that reach $\mathcal{D}_{E}$ before $\Sigma_{0.7}$ are inside, ones that reach $\Sigma_{0.7}$ first are outside. To minimise computational cost, crossings of $\mathcal{D}_{E}$ and $\Sigma_{0.7}$ are monitored during the Lagrangian descriptor calculation. Should the integration time be too short for a significant number of trajectories to reach either surface, we suggest considering these as lying inside the reactive island.
 
\begin{figure}[htbp]
  \begin{center}
    \includegraphics[width=0.35\textwidth]{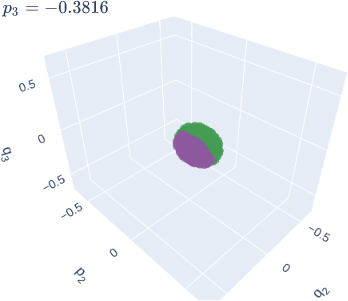}\hfil
    \includegraphics[width=0.35\textwidth]{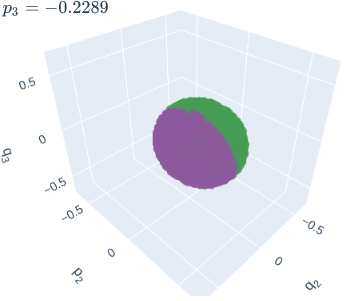}\\
    \includegraphics[width=0.35\textwidth]{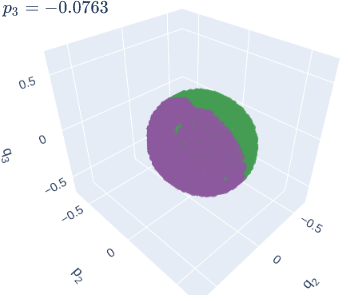}\hfil
    \includegraphics[width=0.35\textwidth]{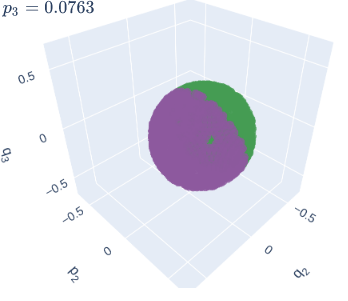}\\
    \includegraphics[width=0.35\textwidth]{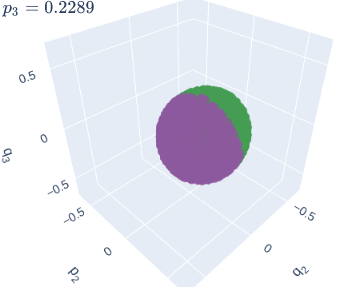}\hfil
    \includegraphics[width=0.35\textwidth]{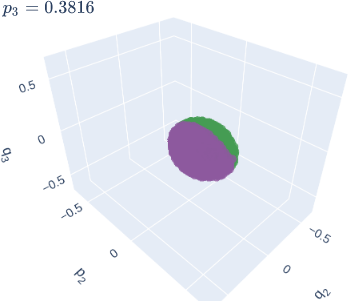}
  \end{center}
  \caption{4D section $q_1=-0.5, p_1>0$ of the intersection of $\mathcal{W}^{s}$ and $\mathcal{W}^{u}$ on the energy surface $E=0.1$ consisting of several $p_3=\text{const}>0$ slices ($p_3<0$ is related by symmetry). The parameters are $\omega_2=\omega_3=1$ and $\varepsilon=0.1$.}
  \label{fig:CDW_intersection}
\end{figure}

We proceed in the same way for trajectories on $\mathcal{W}^{s}$ in backward time. This allows us to identify pieces of $\mathcal{W}^{s}$ and $\mathcal{W}^{u}$ that bound the intersection of the first forward reactive island with the first backward reactive island. We remark that the union of the two reactive islands can be determined analogously. The result is displayed in Fig. \ref{fig:CDW_intersection}. We approximate reactive islands using:
 \begin{equation}
 \int^{\tau}_{0} \Big|\dot{q}_1 - \dfrac{\varepsilon}{c}(\dot{q}_2+\dot{q}_3)\Big| \, dt \,,
 \label{eq:LDcoupled}
 \end{equation}
where $c$ is defined in Eq. \eqref{eq:c}, and a 4-dimensional grid of size $60\times60\times60\times10$ that covers the region
\begin{equation}
 \{|q_2|,|p_2|,|q_3|,|p_3| \leq c_{0.7}\},
\end{equation}
on $\Sigma_{0.7}$, where
\begin{equation}
c_{0.7} = \sqrt{2E-\dfrac{(0.7)^4}{2}+\left(1-2\varepsilon\right) \cdot (0.7)^2}
\end{equation}
is the maximal value attainable by any of the harmonic oscillators for $q_1 = \pm 0.7$. The estimated fluxes are shown in Tab. \ref{tab:coupled_forward} and  \ref{tab:coupled_intersection}, which we compare to a Monte-Carlo estimate of $0.1779$ for the flux across either reactive island and $0.1102$ for their intersection obtained using $10^6$ trajectories launched from $\mathcal{D}_{E}$. It is possible to estimate the fluxes on a coarser grid, we were able to obtain estimates within $\pm10\%$ on a uniform $30\times30\times30\times8$ grid, but we consider these unreliable due to holes in the triangulated surface and deviating Euler characteristics.
  
\begin{table}[h]
   \begin{tabular}{l l|c c c c c}
   \multicolumn{2}{c}{} & \multicolumn{5}{c}{cut-off}\\
    & & 0.6 & 0.8 & 1.0 & 1.1 & 1.2\\
   \hline
   \multirow{3}{*}{\rotatebox{90}{$\tau$}}
    & 5 & 0.1820 & 0.1843 & - & - & -\\
    & 6 & 0.1813 & 0.1803 & 0.1811 & - & -\\
    & 7 & - & 0.1807 & 0.1799 & 0.1797 & -\\
    & 8 & - & - & 0.1814 & 0.1809 & -\\
\end{tabular}
\caption{\label{tab:coupled_forward}Estimated flux across the first forward reactive island in the coupled system ($\varepsilon=0.1$) using LD \eqref{eq:LDcoupled} for selected $\tau$ and cut-off. Estimates for flux across the first backward reactive island are within $10^{-3}$. ``-'' stands for unreliable estimate due to holes in the triangulated surface and deviating Euler characteristics.}
\end{table}

\begin{table}[h]
   \begin{tabular}{l l|c c c c c}
   \multicolumn{2}{c}{} & \multicolumn{5}{c}{cut-off}\\
    & & 0.6 & 0.8 & 1.0 & 1.1 & 1.2\\
   \hline
   \multirow{3}{*}{\rotatebox{90}{$\tau$}}
    & 5 & 0.1147 & 0.1155 & - & - & -\\
    & 6 & 0.1152 & 0.1133 & 0.1144 & - & -\\
    & 7 & - & 0.1139 & 0.1131 & 0.1130& -\\
    & 8 & - & - & 0.1146 & 0.1132 & -\\
\end{tabular}
\caption{\label{tab:coupled_intersection}Estimated flux across the intersection of the first forward and backward reactive islands in the coupled system ($\varepsilon=0.1$) using LD \eqref{eq:LDcoupled} for selected $\tau$ and cut-off. ``-'' stands for unreliable estimate due to holes in the triangulated surface and deviating Euler characteristics.}
\end{table}

When the Poisson reconstruction returns a spherical surface without holes, the agreement with the Monte-Carlo estimate is within $5\%$ and our results are robust under parameter variations. This depends mostly on the density of the grid, as $16$ times as many points (via symmetries) in the uncoupled system returned an accurate results in a broader range of parameters.

On top of the flux estimates we have carried out, our approach allows for a qualitative investigation of the intersection of $\mathcal{W}^{s}$ and $\mathcal{W}^{u}$, which is invaluable for the study of transport and bifurcation analysis.

\section{Conclusions and future work}
\label{conc}

In this paper we have developed the geometrical, analytical, and computational framework for reactive island theory for three degrees-of-freedom time-independent Hamiltonian systems. In the past, reactive island theory has been successfully applied to studies of chemical reaction dynamics from the phase space point of view in 2 DoF systems, including the computation of rates in chemical reactions. This work advances reactive island theory to 3 DoF systems, providing a foundational step towards achieving this goal for high-dimensional systems with 3 or more DoF.

For 2 DoF systems, the dividing surface anchored at the unstable periodic orbit provides the frontier between two dynamically distinct regions (e.g. reactants and products). The 2-dimensional stable and unstable invariant manifolds of the unstable periodic orbit have the geometry of two-dimensional cylinders (or ``tubes'') and they act as phase space conduits that determine whether or not trajectories cross between these dynamically distinct regions. The intersections of these manifolds with an adequately selected isoenergetic two-dimensional Poincar\'e section transverse to the tubes gives rise to the reactive island picture. From this approach, one can calculate the flux of trajectories from one region to the other by means of Stokes theorem.

The geometrical structure of higher-dimensional systems based on normally hyperbolic invariant manifolds and their stable and unstable invariant manifolds has been known for some time, but required a substantial conceptual leap in the computational realisation of these structures. Reactive islands theory in 3 DoF systems is a departure from 2-dimensional Poincar\'e sections and traditional ways of finding stable and unstable invariant manifolds. Instead, recent extensions of the method of Lagrangian descriptors provide the ability to computationally realize and probe these high-dimensional structures using low-dimensional representations. Using these advances, we have shown how this approach can be used to realize a reactive island theory for the benchmark example of a two-well potential energy surface defining a 3 DoF Hamiltonian system.

While the visualisation of the reactive islands and the triangulation procedure we employed only works for 3 DoF systems, these limitations can be overcome to treat higher dimensional systems. Reliance on the visual inspection of high-dimensional structures can and will have to be replaced by numerical approaches, such as the tests of connectedness and Euler characteristic presented in this work. Triangulation procedures for higher dimensional manifolds exist or are being developed by computational geometers. Therefore our framework can be used in higher dimensional systems and only relies on the knowledge of the relevant normally hyperbolic invariant manifolds.

Our work provides an entry point into high-dimensional studies of the geometry of phase space reaction dynamics. Future work might involve the development of kinetic reaction rate models along the lines of earlier work on reactive island theory as well as studies of realistic molecular systems in higher dimensions.

\section*{Acknowledgments}

The authors would like to acknowledge the financial support provided by the EPSRC Grant No. EP/P021123/1 and the Office of Naval Research Grant No. N00014-01-1-0769.

\bibliography{ri_3dof}

\end{document}